\documentclass[pra,amsfonts,showpacs,showkeys,preprint]{revtex4}
\usepackage{xcolor}
\usepackage{graphicx}
\usepackage{dcolumn}
\usepackage{bm}
\usepackage{eepic}
\bibpunct{[}{]}{,}{a}{}{;}
\RequirePackage{times}
\RequirePackage{mathptm}

\begin{document}

\title{Three criteria for quantum random number generators based on beam splitters}

\author{Karl Svozil}
\email{svozil@tuwien.ac.at}
\homepage{http://tph.tuwien.ac.at/~svozil}
\affiliation{Institute for Theoretical Physics, Vienna University of Technology,  \\
Wiedner Hauptstra\ss e 8-10/136, A-1040 Vienna, Austria}

\begin{abstract}
We propose three criteria for the generation of random digital strings from quantum beam splitters:
(i) three or more mutually exclusive outcomes corresponding to the invocation of three- and higher dimensional Hilbert spaces;
(ii) the mandatory use of pure states in conjugated bases for preparation and detection; and
(iii) the use of entangled singlet (unique) states for elimination of bias.
\end{abstract}


\maketitle

Quantum random number generators are important for quantum information
processing, as they are likely to be one of the first technologies applied for various physical and commercial applications.
They also serve as components of other quantum devices for quantum key distribution
and experiments testing and utilizing quantum nonlocality.

Randomness is a notorious property, both from  theoretical and  practical points of view.
It is commonly accepted that there is a satisfactory definition~\cite{calude:02}
of {\em infinite} random sequences in terms of algorithmic incompressibility \cite{ch6}
as well as of the equivalent statistical tests~\cite{MartinLöf1966602}.
Besides the obvious fact that all computable and physically operational entities are limited to {\em finite} objects and methods,
algorithmic pseudo-random generators suffer from von Neumann's verdict that~\cite{von-neumann1}:
{\it ``anyone who considers arithmetical methods of producing random digits is, of course, in a state of sin.''}
The halting probability $\Omega$~\cite{chaitin:01} shares three perplexing properties:
it is computably enumerable (computable in a weak sense), provable random
(which implies that $\Omega$ is non-computable), as well as infinitely knowledgeable in its role as a ``rosetta stone''
for all decision problems encodable as halting problems \cite{calude:02}.
A few of its starting bits have been computed~\cite{calude-dinneen06}, yet due to its randomness,
only finitely many bits of this number can ever be computed.


From the numerous random number generators based on physical processes
(cf. Refs.~\cite{rand-55,0022-3735-3-8-303,er-put:85,schmidt:462,Martino:91,walker-hotbits} to name but a few),
the use of single photons (or other quanta such as neutrons) subjected to beam splitters
appears particularly promising~\cite{svozil-qct,rarity-94,zeilinger:qct,stefanov-2000} for the following reasons:
(i) due to (ideally) single-photon events, the physical systems are ``elementary;''
(ii) they can be controlled to a great degree; and
(iii) they can be certified to be random relative to the postulates of quantum theory~\cite{zeil-05_nature_ofQuantum}.

Three features of quantum theory directly relate to random sequences generated from beam splitter experiments:
(i) the randomness of individual events (cf. Ref.~\cite[p.~866]{born-26-1} and Ref.~\cite[p.~804]{born-26-2});
(ii) complementarity~\cite[p.~7]{pauli:58}; and
(iii) value indefiniteness; i.e., the absence of two-valued states
interpretable as ``global'' (i.e., valid on all observables) truth functions~\cite{kochen1}.
In order to fully implement these quantum features,
we propose three improvements to existing protocols~\cite{svozil-qct,rarity-94,zeilinger:qct,stefanov-2000,0256-307X-21-10-027,wang:056107,fiorentino:032334}.

The first criterion ensures that the quantum random number generators can be certified to be subjected to quantum value indefiniteness.
A necessary condition for this to apply is the possibility
of {\em three or more mutually exclusive outcomes} in measurements of single quanta.
Formally, this is due to the fact that violations of Bell-type inequalities, as well as proofs of Gleason's
and  Kochen-Specker-type theorems are only realizable~\cite{svozil-tkadlec} in three- and higher dimensional Hilbert spaces.
Only from three-dimensional vector space onwards it is possible to nontrivially interconnect bases through
one  (or up to $n-2$ for $n$-dimensional Hilbert space) common base element(s).
This can be explicitly demonstrated by
certain, even dense~\cite{godsil-zaks,meyer:99,havlicek-2000}, ``dilutions'' of bases, which break up the possibility to interconnect,
thus allowing value definiteness.
In more operational terms, if some ``exotic'' scenarios (e.g., Ref.~\cite{pitowsky-82,pitowsky-83}) are excluded,
the violation of Bell-type inequalities for two two-state particles (corresponding to two outcomes on each side)
is a sufficient criterion for quantum value indefiniteness.

Of course, one could  argue that protocols based on two outcomes are still protected by quantum complementarity,
and the full range of quantum indeterminism, in particular quantum value indefiniteness, is not needed.
There is also the possibility that the Born rule might be derived through some other argument
(possibly from another set of axioms)
than Gleason's theorem~\cite{Gleason,pitowsky:218,rich-bridge,r:dvur-93}.
However, there exist sufficiently many two-valued states on propositional structures with two outcomes
to allow for a homeomorphic embedding of this structure into a classical Boolean algebra.
In any case, it appears prudent to use all the ``mind-boggling'' features of quantum mechanics against
cryptananalytic attacks on some quantum-generated sequence.

The resulting trivalent or multi-valued sequence can be easily ``downgraded'' or ``translated''
to binary sequences through elimination or identification without loss of randomness:  systematically eliminating $n-2$ symbol(s)
will transform a random sequence on an alphabet with $n \ge 3$  symbols into a random sequence on an alphabet with two
symbols \cite{calude:02}.

\if01
if the sequence contains an even number of symbols, the set of symbols could be arbitrarily equipartitioned into two parts,
followed by the identification of the first and the second partition element with ``$0$'' and ``$1$,'' respectively.
For example, if there are four  mutually exclusive outcomes encoded by the symbols $1,2,3,4$, then any one of the equipartitions
$\{\{1,2\},\{3,4\}\}$,
$\{\{1,3\},\{2,4\}\}$, and
$\{\{1,4\},\{2,3\}\}$ can be used for this two-to-one mapping; e.g., $\{1,2\} \mapsto 0$  and  $\{3,4\} \mapsto 1$.
If the sequence contains an even number of symbols, then it suffices to eliminate one of these symbols from the sequence~\cite{MR997340}
and apply the previous procedure for sequences with an even number of symbols.
\fi

The second criterion proposes the mandatory use of {\em pure} states from maximally conjugated bases for preparation and detection.
This requirement deals with the {\em single particle source} of quantum random number generators.
Indeed, many two-particle experiments have been using this criterion already, as
full tomography is performed to characterize the state as completely as possible.
These experiments use a (Bell) state which is as pure and maximally entangled as operationally feasible; quite
often they produce the singlet Bell state (which, due to technical issues related to other degrees of freedom,
can never be ideally pure). Tomography is used to characterize the state and hence certify the randomness of outcomes.
Hence in this sense and in these experiments, the criterion is already implicitly implemented.

Although it is generally believed that mixed (nonpure) quantum states can be ``produced''
and operationalized ``for all practical purposes,''
one might cautiously argue that this may actually be a subjective statement on behalf of the observer:
whereas the experimenter might ``pretend'' that the exact state leaving the particle source is unknown,
it might still be possible to conceive of the state to be in some, albeit unknown but not principally unknowable, unique pure state.
This is related to the question of whether or not mixed states should be thought of as merely subjective constructions
which even in the epistemic view --- as the wave function (the quantum state) representing a catalog of expectations~\cite{schrodinger}
--- represent only certain partial, incomplete representations of systems which might be completely defined by a single unique context.

Even if one is unwilling to accept these principal concerns, it remains prudent
not to expose the protocols for generating quantum randomness
to the possibility of hidden regularities of the source.
After all, beam splitters are just one-to-one bijective devices representable by reversible
unitary operators~\cite{Mandel-Ou1987118,green-horn-zei,zeilinger:882}; a fact which can be seen by recombining the two paths by a second beam splitter in a Mach-Zender
interferometer, thereby recovering the original signal.
Thus, in order to assure quantum randomness, the beam splitter should not be considered as an isolated element,
but has to be examined in combination with the source.
In accordance with this principle, a {\em mismatch} between state preparation and  measurement guarantees
that quantum complementarity ensures the indeterministic outcome.
This can, for instance, be implemented by preparing the single particle in a pure state which corresponds to an element of a certain basis,
and then measuring it in a different basis,
in which the original state is in a coherent superposition of more than one states
(cf. Ref.~\cite{svozil-qct} and the first protocol using beam splitting polarizers in Ref.~\cite{zeilinger:qct}).

Third and finally,  in order to eliminate any possible bias
(for some ``classical'' methods to eliminate bias, we refer to Refs.~\cite{elias-72,PeresY-1992,dichtl-2007,Lacharme-2008}),
we propose  to utilize
Einstein-Podolsky-Rosen type  measurements  of two quanta in a unique entangled state.
Any state satisfying the uniqueness property~\cite{svozil-2006-uniquenessprinciple} in at least two directions,  such as the singlet states
${1\over \sqrt{2}} \left(  \vert \frac{1}{2},-\frac{1}{2} \rangle - \vert -\frac{1}{2},\frac{1}{2} \rangle \right)$,
$ \frac{1}{\sqrt{3}}\left(-|0,0\rangle + |-1,1\rangle + |1,-1\rangle \right)$, or
$\frac{1}{2} \left(
\left| \left. \frac{3}{2}, -\frac{3}{2}\right\rangle \right.
 - \left| \left.  -\frac{3}{2}, \frac{3}{2}\right\rangle    \right.
- \left| \left.  \frac{1}{2}, -\frac{1}{2}\right\rangle  \right.
+ \left| \left.  -\frac{1}{2}, \frac{1}{2}\right\rangle   \right.
\right)$ of two spin--$\frac{1}{2}$, -$1$, or -$\frac{3}{2}$ particles, could be used for this purpose.
In that way, the outcome of one particle can be combined with the outcome of the other particle to eliminate bias.
Again, it should be kept in mind that physical realizations of this protocol can never be made ideal
and necessarily suffer from, for instance, the nonideal behavior of the beam splitters.

For the sake of demonstration, suppose Alice and Bob share successive pairs of quanta
in the singlet Bell state ${1\over \sqrt{2}} \left(  \vert \frac{1}{2},-\frac{1}{2} \rangle - \vert -\frac{1}{2},\frac{1}{2} \rangle \right)$.
Denote Alices's and Bob's outcomes in the $j$th measurement by $a_j$ and $b_j$,
with the coding $a_j,b_j \in \{0,1\}$,
respectively.
Using XOR operations on their combined results by a product mod~2 of $a_j$ and $b_j$, i.e., by defining
$s_j = a_j \oplus b_j = a_j  b_j \text{ mod }2$,
yields a totally unbiased sequence $s_j$ of bits.
Remarkably, as the state guarantees a 50:50 occurrence of 0's and 1's on either side,
the associated bases of Alice and Bob  need not even be maximally ``apart'':
one outcome on Alice's side can be thought of as serving as ``one time pad'' in encrypting the
other outcome on Bob's side, and vice versa.
Again, this method will be as good as the entangled particle source.
In order to eliminate causal influences, the events recorded by Alice and Bob should be
separated by strict {E}instein locality conditions \cite{wjswz-98,arXiv:0811.3129},
although separating the particles will be experimentally challenging.

Alternatively, in an adaptive ``delayed choice'' experiment
the outcome on Alice's side could be transferred to Bob, who adjusts his experiment (e.g., by changing the direction of spin state measurements)
according to Alice's input~\cite{VincentJacques02162007}.
This method resembles the previously implemented self-calibration techniques
utilizing coincidence measurements~\cite{0256-307X-21-10-027},
entropy measures~\cite{fiorentino:032334},
and iterative sampling~\cite{wang:056107}.
Whether or not it could also be used for classical angular momentum zero states ``exploding''
into two parts~\cite{peres222} remains unknown.

In summary we have argued that the present protocols for generating quantum random sequences with beam splitters should be improved
to be certifiable against value definiteness and hidden bias of the source.
We have also proposed a procedure to eliminate bias by using one particle of a singlet in an Einstein-Podolsky-Rosen configuration
as a one-time pad for the other particle.

$\;$\\
{\bf Acknowledgements}
\\
The author gratefully acknowledges discussions with and suggestions by Cristian Calude, as well as the kind hospitality of the {\it Centre for Discrete Mathematics
and Theoretical Computer Science (CDMTCS)} of the {\it Department of Computer Science at
The University of Auckland.}
This work was also supported by {\it The Department for International Relations}
of the {\em Vienna University of Technology.}


\end{document}